\documentclass{JHEP3}
\usepackage{amssymb}

\newcommand{\fr}{\frac}
\newcommand{\lb}{\label}
\newcommand{\ti}{\tilde}

\newcommand{\be}{\begin{equation}}
\newcommand{\ee}{\end{equation}}
\newcommand{\ba}{\begin{array}}
\newcommand{\ea}{\end{array}}
\newcommand{\beqa}{\begin{eqnarray}}
\newcommand{\beqas}{\begin{eqnarray*}}
\newcommand{\eeqa}{\end{eqnarray}}
\newcommand{\eeqas}{\end{eqnarray*}}
\newcommand{\al}{\alpha}
\newcommand{\bt}{\beta}
\newcommand{\ka}{\kappa}
\newcommand{\la}{\lambda}
\newcommand{\La}{\Lambda}

\newcommand{\si}{\sigma}
\newcommand{\ro}{\rho}
\newcommand{\te}{\theta}
\newcommand{\Te}{\Theta}
\newcommand{\del}{\partial}
\newcommand{\ep}{\epsilon}

\newcommand{\kd}{\delta}

\newcommand{\vae}{\varepsilon}
\def\dels{\partial\!\!\! /}
\def\delsbar{\bar{\partial\!\!\! /}}

\def\a{\alpha}
\def\b{\beta}

\def\e{\epsilon}

\def\t{\theta}
\def\k{\kappa}
\def\la{\lambda}
\def\t{\theta}
\def\s{\sigma}

\def\p{\psi}

\def\adot{\dot{\alpha}}
\def\bdot{\dot{\beta}}

\def\labar{\bar{\lambda}}
\def\sbar{\bar{\sigma}}
\def\pbar{\bar{\psi}}

\def\tbar{\bar{\theta}}

\def\intx{\int d^4 x}

\title{Duals of noncommutative supersymmetric
 $\mathbf U(1)$ gauge theory} 

\author{\"{O}mer F. Dayi \\
 Physics Department, Faculty of Science and
Letters, Istanbul Technical University,
 80626 Maslak--Istanbul,
Turkey, and\\
 Feza G\"{u}rsey Institute,
 P.O.Box 6, 81220
\c{C}engelk\"{o}y--Istanbul, Turkey \\
E-mail: \email{ dayi@itu.edu.tr,  dayi@gursey.gov.tr} }
\author{Kayhan \"{U}lker \\
 Feza G\"{u}rsey Institute,
 P.O.Box 6, 81220
\c{C}engelk\"{o}y--Istanbul, Turkey \\
E-mail: \email{kulker@gursey.gov.tr} }
\author{Bar\i\c{s} Yap\i\c{s}kan \\
 Physics Department, Faculty of Science and
Letters, Istanbul Technical University,
 80626 Maslak--Istanbul,
Turkey\\
E-mail: \email{yapiska1@itu.edu.tr}}

\abstract{Parent actions for 
component fields are utilized to derive
the dual of supersymmetric  $U(1)$ gauge theory in 4 dimensions. 
Generalization of the Seiberg--Witten map to the component fields of
noncommutative supersymmetric $U(1)$ gauge
theory is analyzed. 
Through this transformation we proposed  parent actions
for noncommutative supersymmetric  $U(1)$ gauge theory
as generalization of the ordinary case.
Duals of noncommutative
supersymmetric  $U(1)$ gauge theory
are obtained. Duality symmetry under the interchange 
of fields with duals 
accompanied by
the replacement of the noncommutativity parameter
$\Theta_{\mu \nu}$ with 
$\tilde{\Theta}_{\mu \nu} = 
g^2\epsilon_{\mu\nu\rho\sigma}\Theta^{\rho\sigma}$
of the non--supersymmetric case is broken at the level of actions.
We proposed a noncommutative parent action
for the component fields
which generates actions possessing this duality symmetry.}


\preprint{\hepth{0309073}}

\begin{document}

\section{Introduction}

Electric--magnetic duality invariance of Maxwell equations
can be formulated at the level of actions due to a parent
action which generates both the original and the dual actions.
This approach was used to derive dual of noncommutative
 $U(1)$ gauge  theory\cite{grs} after transforming noncommutative
gauge theory to commutative one by the Seiberg--Witten 
map\cite{sw}. Here we 
study  dual transformations
of noncommutative supersymmetric  $U(1)$ gauge theory
in 4 dimensions using a 
similar procedure  considering  the component fields of
superfields. 

Parent action of ``ordinary" supersymmetric $U(1)$ gauge theory
was formulated  by superfields\cite{sw0}. In terms of
 component fields we define two different parent
actions which yield the same dual symmetric actions.

Noncommutative supersymmetric $U(1)$ 
gauge theory can be defined
as generalization of  supersymmetric 
Yang--Mills gauge theory either
through superfields\cite{fl} 
or using their component fields\cite{ter}.
To derive its dual theory by performing duality
transformation for the ordinary fields as
in \cite{grs}, one should find a transformation 
which generalizes
the Seiberg--Witten map\cite{sw} to noncommutative
supersymmetric $U(1)$ gauge theory fields.
This transformation was studied in two different ways through
superfields\cite{fl},\cite{ck}. We utilize both of
these approaches to define a generalization of
the Seiberg--Witten map for the component fields.
Then we write noncommutative supersymmetric
$U(1)$ gauge theory in terms of the component
fields which are valued in commuting space--time.
We only deal with the terms up to the first
order in the noncommutativity parameter $\Te_{\mu\nu} .$

Generalizing parent actions of ordinary supersymmetric 
gauge theory and using the map between  
``noncommutative" and ordinary (commutative)
fields we propose two different parent actions.
Both of them generate noncommutative 
supersymmetric $U(1)$ gauge theory given by the component
fields defined in commuting space--time.
However, they  yield different dual actions.
At the first order in $\Te_{\mu\nu} $
one of the dual actions does not have any contribution from
the fermionic and the auxiliary  fields. Moreover, 
it does not lead to the dual action 
of non--supersymmetric gauge theory
of \cite{grs}.
The other parent action generates a dual theory which
embraces the results of \cite{grs}. However, 
this dual action is not in the same
form with the noncommutative $U(1)$ gauge theory. Thus,
duality symmetry of the non--supersymmetric theory  given
by replacing the field strength $F^{\mu\nu}$ with
the dual one $F_{D}^{\mu\nu}$ and $\Te_{\mu \nu}$
with $\ti{\Te}_{\mu\nu}=g^2\ep_{\mu\nu\rho\sigma}\Te^{\rho \sigma}$
is not satisfied
when actions are considered.
We introduce a parent action for the
component fields  which generates 
actions possessing this duality symmetry. Unfortunately,
it is not clear if these duality symmetric actions
are supersymmetric, though they are explicitly
gauge invariant.

\section{Parent actions for component fields}

Working in 4 dimensional Minkowski space--time and the $N=1$
superspace $(x_\mu , \te_\al , \bar{\te}^{\adot})$ we consider 
a general chiral superfield (not a supersymmetric field strength) 
$\ti{W}_{\al}$ 
and a real (dual) vector field $V_D$ to write the parent
action\cite{sw0}
\be
\lb{sfd}
I_{p}=\fr{1}{4g^2}  \int d^{4}x( \int d^{2}\te\ti{W}^{2} + \int d^{2}
\bar\te {\bar{\ti{W}}}^{2})+\fr{1}{2} \int d^{4}xd^{4}\te
(V_{D}D\ti{W}-V_{D}\bar{D}\bar{\ti{W}})
\ee
where 
$D_\al$ is the supercovariant derivative. We use
notations of \cite{wb}.
Equation of motion
with respect to the  super vector field $V_D$ 
leads to the supersymmetric generalization of the
Bianchi identity
\be
\lb{real}
D\ti{W} - \bar{D} \bar{\ti{W}}|_W= 0.
\ee
Its solution  
is the supersymmetric field strength
written in terms of the
real  vector superfield $V$ as 
\be
\lb{sftg}
W_\al = \fr{1}{2}{\bar{D}}^2 D_\al V.
\ee

Replacement of $\tilde{W},\bar{\tilde{W}}$ 
with the  solution (\ref{sftg})
in the parent action (\ref{sfd}),
which is equivalent
to perform the path integral over $V_D$ in its 
partition function, leads to
\be
\lb{aosu}
I=\fr{1}{4g^2}  \int d^{4}x (\int d^{2}\te W^{2} +
\int d^{2}\bar{\te}\bar{W}^{2}).
\ee
This is the action of 
supersymmetric U(1) gauge theory.

On the other hand, when 
solutions of 
the equations of motion with respect to 
$\ti{W}_{\al}$
and  $\bar{\ti{W}}^{\dot{\al}}$
following from  $I_p$
are plugged into (\ref{sfd}), one obtains
the dual action 
\be
\lb{i1d}
I_{D}=
\fr{g^2}{4}  \int d^{4}x (\int d^{2}\te W^{2}_D +
\int d^{2}\bar{\te}\bar{W}^{2}_D)
\ee
where  $W_D$ is the dual superfield strength
$W_{D\al} = 
\fr{1}{2}{\bar{D}}^2 
D_\al V_D$.

The original and the dual actions,  (\ref{aosu})
and  (\ref{i1d}),
are in the same form except $g^{-2}$ replaced with  $g^2.$
Thus, one can conclude that
supersymmetric $U(1)$ gauge theory
possesses (S) duality symmetry.

Instead of superfields,
we would like to consider 
duality transformations 
 in terms of their
component fields.
It is straightforward to construct a general chiral 
superfield $\ti W_\al$ that  does not satisfy
 the condition (\ref{real}) as
\be
\lb{gsup}
\ti W_\a (y) = -i\la_\a (y) +\t_\a \ti{D}(y) - i\s^{\mu\nu\,\b}_\a 
\t_\b 
\ti{F}_{\mu\nu}(y)
+ \t\t \s_{\a\adot}^\mu \del_\mu \pbar^{\adot}(y)
\ee
where $y_\mu = x_\mu +i \t\s_\mu\tbar $. Here, $\la$ and
$\pbar$ are two independent Weyl spinors, 
$\ti{F}_{\mu\nu} $ is a complex 
antisymmetric  field and $\ti{D}$ is a complex
scalar field. Hermitean conjugate of the chiral
superfield $\ti{W}_\a $ can be written as
\be
\lb{gsud}
\bar{\tilde{W}}^{\adot}  (y^\dag) = i\labar^{\adot} (y^\dag) +\tbar^{\adot} 
D^\dag (y^\dag)
+ i\sbar^{\mu\nu\adot} _{ \,\bdot }\tbar^{\bdot} F_{\mu\nu}^\dag (y^\dag)
+ \tbar\tbar \sbar^{\adot\a}_\mu \del^\mu \p_{\a}(y^\dag).
\ee
Plugging (\ref{gsup}), (\ref{gsud}) and 
the real  vector superfield
\be
V_D=-(\te\si^{\mu}\bar{\te})A_{D\mu}
+ i \te\te\bar{\te}\bar{\la}_D
- i \bar{\te}\bar{\te}\te\la_D
+ \fr{1}{2}\te\te\bar{\te}\bar{\te}D_D
\ee
into (\ref{sfd})   
the parent action in component fields is obtained
\be
\lb{sfdc}
I_p=I_o 
[\ti{F},\psi, \la, \ti{D}]
+I_l 
\ee  
where we defined
\beqa
I_o[\ti{F},\psi, \la, \ti{D}]
 &\equiv& \frac{1}{g^2}\intx [- \frac{1}{8}\ti{F}^{\mu\nu}\ti{F}_{\mu\nu} 
- \frac{i}{16}\e^{\mu\nu\la\k}\ti{F}_{\mu\nu}\ti{F}_{\la\k} 
- \frac{1}{8}\ti{F}^{\dag\mu\nu}\ti{F}^\dag _{\mu\nu} \nonumber\\ 
& & + \frac{i}{16}\e^{\mu\nu\la\k}\ti{F}^\dag _{\mu\nu}
\ti{F}^\dag _{\la\k}
  -\frac{i}{2}\la \dels \pbar -
\frac{i}{2}\labar \delsbar \p 
+\frac{1}{4}\ti{D}^{2} +\frac{1}{4}\ti{D}^{\dag 2} ] \lb{iod}
\eeqa
and the Legendre transformation term
\beqa
I_l & \equiv &\fr{1}{2}\intx [ -i\ti{F}^{\mu\nu}\del_\mu A_{D\nu}
+  \frac{1}{2}\e^{\mu\nu\la\k}\ti{F}_{\mu\nu}\del_\la A_{D\k} 
+i\ti{F}^{\dag\mu\nu}\del_\mu A_{D\nu} \nonumber\\
& & +  \frac{1}{2}\e^{\mu\nu\la\k}\ti{F}^\dag_{\mu\nu}
\del_\la A_{D\k} +\frac{1}{2}\la_D \dels \pbar 
+\la \dels \labar_D - \frac{1}{2}\labar_D \delsbar \p
 -\labar \delsbar \la_D 
+iD_D (\ti{D} -\ti{D}^\dag )] . \lb{ild}
\eeqa

We now proceed as before to derive 
supersymmetric $U(1)$ gauge theory 
in terms of the component fields
from the parent action 
(\ref{sfdc}):
The equations of motion with respect to
the dual vector field $A_{D\mu}$ 
\be
\lb{hdd1}\left[
\frac{i}{2}(\del_\mu \ti{F}^{\mu\k} - \del_\mu \ti{F}^{\dag\mu\k}) 
-\frac{1}{4} \e^{\mu\nu\la\k}\del_\la (\ti{F}_{\mu\nu} + 
\ti{F}^\dag _{\mu\nu})\right]_{\ti{F}=F} =0.
\ee
lead to $F_{\mu\nu}$ which satisfy
\be
F_{\mu\nu}=F^\dag _{\mu\nu}, \, ,\,\, \e^{\mu\nu\la\k}\del_\la 
F_{\mu\nu}=0 ,
\ee
which are solved by taking
$F_{\mu \nu}=\del_\mu A_\nu-\del_\nu A_\mu $
which is the field strength of the vector field $A_\mu.$
When we also use  
the equations of motion with respect to  the other dual fields 
\be
\lb{spsi}
\dels \pbar = \dels \labar\,\, ,\,\, \delsbar \p =\delsbar 
\la \,\, ,\,\, \ti{D}-\ti{D}^\dag|_{\ti{D}=D}=0 .
\ee
in the parent action (\ref{sfdc}) we obtain 
the supersymmetric   $U(1)$ gauge theory action 
in terms of component fields
\be
\lb{smt}
I =  \frac{1}{g^2}\int d^4x [- \frac{1}{4}F^{\mu\nu}F_{\mu\nu} 
-\frac{i}{2}\la \dels \labar -\frac{i}{2}\labar \delsbar \la 
+\frac{1}{2}D^{2} ].
\ee

Similarly, we can obtain the dual action (\ref{i1d}) in terms of
the component fields using 
the equations of motion of (\ref{sfdc}) with respect to the fields 
$\ti{F}_{\mu\nu},\la ,\pbar ,\ti{D}:$ 
\beqa
(\eta^{\mu\la}\eta^{\nu\k}- \eta^{\mu\k}\eta^{\nu\la}
+i\e^{\mu\nu\la\k})F_{\la\k}
= - ig^2 (\eta^{\mu\la}\eta^{\nu\k}- \eta^{\mu\k}\eta^{\nu\la}
+i\e^{\mu\nu\la\k})F_{D\la\k}  , \lb{hdo1}\\
\dels \pbar = -ig^2\dels \labar_D \,\, ,\,\,\delsbar 
\la = -ig^2\delsbar \la_D \,\, ,\,\,  
\ti{D}= -ig^2 D_D
\eeqa
and the equations of motion  with respect to $\ti{F}^\dag _{\mu\nu},\labar 
,\p , \ti{D}^\dag :$
\beqa
(\eta^{\mu\la}\eta^{\nu\k}- \eta^{\mu\k}\eta^{\nu\la}-i\e^{\mu\nu\la\k})
\ti{F}^\dag _{\la\k}
=  ig^2 (\eta^{\mu\la}\eta^{\nu\k}- 
\eta^{\mu\k}\eta^{\nu\la}-i\e^{\mu\nu\la\k})
F_{D\la\k} ,\\
\dels \labar = ig^2\dels \labar_D \,\, ,\,\,\delsbar \p 
= ig^2\delsbar \la_D \,\, ,\,\,  
\ti{D}^\dag = ig^2 D_D , \lb{hdo4}
\eeqa
where $F_{D\mu\nu}= \del_\mu A_{D\nu} - \del_\nu A_{D\mu}.$
Solutions of them 
are plugged  into (\ref{sfdc}) yielding
the  dual supersymmetric $U(1)$ gauge theory action
(\ref{i1d}) in terms of the component fields
\be
\lb{smtd}
I_{D} = g^2 \intx [- \frac{1}{4}F_D ^{\mu\nu}F_{D\mu\nu} -\frac{i}{2}\la_D 
\dels \labar_D -\frac{i}{2}\labar_D \delsbar \la_D +\frac{1}{2}D_D ^{2} ] .
\ee

Instead of the complex  field
 $\ti{F}_{\mu\nu}$  we can deal with the real 
antisymmetric tensor field  $F_{R\mu\nu}$ from the 
beginning. For this case 
we propose 
\be
\lb{pr0}
S_{p}=S_o[F_R,\psi,\la,\ti{D}] +S_l,
\ee
as parent action,
where
\be
\lb{so}
S_o[F_R,\psi,\la,\ti{D}] \equiv \frac{1}{4g^2}\int d^{4}x [ - 
F_R^{\mu\nu}F_{R\mu\nu}
-2i\bar{\la}\si^{\mu}\del_\mu \psi
- 2i\la\si^{\mu} \del_\mu \bar{\psi}
+\ti{D}^2 +  \ti{D}^{\dag 2} ] , 
\ee
and the Legendre transformation part
\beqa
S_l & \equiv & \fr{1}{2}\int d^{4}x [ \vae^{\mu\nu\ro\si}
F_{R\mu\nu}\del_\rho A_{D\si}+ \la_{D}\si^{\mu}\del_\mu \bar{\psi}
+  \bar{\la}_{D}\bar{\si}^{\mu}\del_{\mu}\la \nonumber \\
&& -\la_{D}\si^{\mu}\del_{\mu}\bar{\la}-  \bar{\la}_{D}
\bar{\si}^{\mu}\del_\mu \psi + iD_{D}(\ti{D}-\ti{D}^\dag )] \lb{sl} .
\eeqa
The  equations of motions with respect to the dual fields
$A_D,\  \la_D,\  \bar{\la}_D,\ D_D,$
\beqa
\vae^{\mu\nu\ro\si}\del_{\nu}F_{R\ro\si}|_{F_R=F} & = &0 ,\lb{m1}\\
\si^{\mu}_{\al\dot{\al}} \del_\mu \psi^{\dot{\al}}-
\si^{\mu}_{\al\dot{\al}}\del_{\mu}\bar{\la}^{\dot{\al}}&=&0, \\
\bar{\si}_{\mu}^{\dot{\al}\al}\del^{\mu}\la_{\al}
- \bar{\si}_\mu^{\dot{\al}\al} \del^\mu \psi_\al&=&0 ,\\
\left( \ti{D}-\ti{D}^\dagger\right)_{\ti{D}=D }&=&0, \lb{m4}
\eeqa
are solved in terms of the field strength $F_{\mu\nu}$
and real scalar field D. 
These solutions when 
used in the parent action (\ref{pr0})  yield
the supersymmetric  $U(1)$ gauge theory (\ref{smt}).

The equations of motions with respect to the  fields 
$F_{R\mu\nu}, \la, \psi, \bar{\la} , \ti{D},
\bar{\psi}, \ti{D}^\dag $ are
\begin{eqnarray}
  -\fr{1}{g^{2}}F_R^{\mu\nu} +  \vae^{\mu\nu\ro\si}\del_{\ro}A_{D\si}
=  0  & &  \\
  \fr{1}{g^{2}}\ti{D}^\dag - i D_{D} = 0, & &
 \fr{1}{g^{2}} \ti{D}  + i D_{D} = 0 , \nonumber \\
 \si^\mu_{\al\dot{\al}} \del_\mu \left(
-\fr{i}{g^{2}}
\bar{\psi}^{\dot{\al}} + 
\bar{\la}_{D}^{\dot{\al}}\right) = 0 , &  &
\bar{\si}^{\mu\dot{\al}\al}\del_\mu
\left( - \fr{i}{g^{2}} 
\psi_\al
- \la_{D\al}\right) = 0,  \nonumber \\
  \del_\mu (-\fr{i}{g^{2}}
\bar{\la}_{\dot{\al}} 
+\bar{\la}_{D\dot{\al}})\bar{\si}^{\mu\dot{\al}\al}= 0, & &
\del_\mu (- \fr{i}{g^{2}}
\la^{\al}
+ \la^{\al}_{D}) \si^{\mu}_{\al\dot{\al}} = 0 . \nonumber 
\end{eqnarray}
Solving these equations for the dual fields and
substituting them  in the parent action 
(\ref{pr0}) yield the dual of action of N=1 supersymmetric  U(1) gauge theory 
(\ref{smtd}).

We conclude that both of the  parent actions (\ref{sfdc}) and (\ref{pr0})
generate
supersymmetric $U(1)$ gauge theory and its dual.

\section{Supersymmetric  Seiberg--Witten map}

Noncommutativity is introduced
through the star product
\be
\lb{star}
\ast\equiv\exp \frac{i\te^{\mu\nu}}{2} \Big(
{\stackrel\leftarrow\del}_{\mu} {\stackrel\rightarrow\del}_{\nu}
-{\stackrel\leftarrow\del}_{\nu}{\stackrel\rightarrow\del}_{\mu}
\Big),
\ee
where $x_\mu$ are  space--time coordinates and 
$\te^{\mu \nu}$ is an antisymmetric and  constant real parameter.
Now, the coordinates $x^\mu$ satisfy the Moyal bracket
\be
\lb{nc}
x^\mu\ast x^\nu -
x^\nu\ast x^\mu =i\te^{\mu \nu}.
\ee          
We assume that surface terms are vanishing, so that 
the following properties are satisfied
\[
\int d^4xf(x)\ast g(x)=\int d^4xf(x)g(x),  
\]
\[
\int d^4xf(x)\ast g(x)\ast h(x) = \int d^4x(f(x)\ast g(x))h(x)=
\int d^4x f(x)(g(x)\ast h(x)).  
\]

Generalization of  the Seiberg-Witten map to 
noncommutative supersymmetric gauge
theories can be formulated in some different ways.
One of these is to  generalize the 
definition of the map between the
noncommutative gauge field $ \widehat{A},$
noncommutative gauge parameter
$\widehat{\la}$ and 
the ordinary ones
$A,\la $ to $ \widehat{V}(V),\widehat{\La}(\La ,V). $  
Here
V is a vector superfield, $\La$ is a chiral superfield
and $\widehat{V}$
and $\widehat{\La}$ are corresponding ``noncommutative
superfields"\cite{ck}. Infinitesimal gauge transformation 
of the noncommutative supervector field $\widehat{V}$
is defined by
\be
\lb{ncsg}
\widehat{\kd}_{\widehat{\La}}\widehat{V}
=i(\widehat{\La} -\widehat{\bar{\La}})
-\fr{i}{2}[
(\widehat{\La} +\widehat{\bar{\La}})\ast \widehat{V}
-\widehat{V} \ast
 (\widehat{\La} +\widehat{\bar{\La}})] .
\ee
It has the properties of a non-abelian gauge
transformation, although
the ordinary vector field $V$
gauge transforms as
\be
\lb{orsg}
\kd_\La V=i(\La -\bar{\La}) .
\ee

Supersymmetric Seiberg--Witten map 
is defined as
\be
\label{def1}
\widehat{V}(V)+\widehat{\kd}_{\widehat{\La}}\widehat{V}(V)=\widehat{V}(V+\kd_\La V).
\ee
In \cite{ck} a solution of this equation is given in terms
of superfields. However, it is nonlocal and do not yield
the original  solution of Seiberg and Witten\cite{sw}
\be
\label{sws}
\widehat{A}_{\mu}=A_\mu  -\fr{1}{2}\Te^{kl}(A_{k}\del_{l}
A_{\mu}+A_{k}F_{l\mu}) ,
\ee
at the first order in the noncommutativity parameter $\Te^{\mu\nu}.$

 On the other hand the approach suggested in \cite{fl} is to
generalize the solution of Seiberg and Witten  (\ref{sws}) 
to supersymmetric case as
\beqa
\widehat{V}(V) &=& \fr{1}{2}V + a \Te^{\mu\nu} 
\del_{\mu} 
\si_\nu^{\al\dot{\al}} [D_\al ,D_{\dot{\al}}] V
+b 
\si_{\mu\nu}^{\al\bt}\Te^{\mu\nu}
D_{\al}V  {\bar{D}}^2D_\bt V \nonumber \\
 & & + c
\si_{\mu\nu}^{\al\bt}\Te^{\mu\nu}
V  {\bar{D}}^2D_\al D_\bt V 
+ c.c., \\
\widehat{\La}(\La,V) & = & \La + d {\bar{D}}^2\left(
\si_{\mu\nu}^{\al\bt}\Te_{\mu\nu}
D_{\al} D_{\bt}V\right) ,
\eeqa
where a,b,c,d are some constants which
should be fixed using (\ref{sws}). 
Though, it can be solved directly it is cumbersome.
Indeed, its solution is not presented in \cite{fl}.

We would like to obtain a generalization of 
the Seiberg--Witten map to
supersymmetric $U(1)$ gauge theory in terms of the
components of the superfield $V.$ This will be performed 
utilizing both of the methods mentioned above.
We adopt the definition (\ref{def1}) for supersymmetric 
Seiberg--Witten map
but solve it for components of the superfield $V$
 keeping the original solution (\ref{sws}). 

The vector superfield $V$ in Wess-Zumino gauge and chiral and
anti-chiral superfields  $\La $ and $\bar{\La}$ , respectively,
are given as
\begin{eqnarray}
V&=&-(\te\si^{\mu}\bar{\te})A_{\mu}
+ i \te\te\bar{\te}\bar{\la}
- i \bar{\te}\bar{\te}\te\la
+ \fr{1}{2}\te\te\bar{\te}\bar{\te}D, \lb{fwz}\\ 
\La&=&\beta + i (\te\si^{\mu}\bar{\te})\del_\mu\beta +
\fr{1}{4}\te\te\bar{\te}\bar{\te} \del^{2} \beta
+ \sqrt{2} \te\kappa -
\fr{i}{\sqrt{2}}\te\te\del_{\mu}\kappa\si^{\mu}\bar
{\te} + \te\te f ,\lb{Vx}\\
\bar{\La}&=&\beta^{\ast}
- i (\te\si^{\mu}\bar{\te})\del_\mu\beta^{\ast}
+ \fr{1}{4}\te\te\bar{\te}\bar{\te} \del^{2}
\beta^{\ast}+\sqrt{2}\bar{\te}\bar{\kappa}
+\fr{i}{\sqrt{2}}\bar{\te}\bar{\te}\te\si^{\mu}\del_
{\mu}\bar{\kappa}
+ \bar{\te} \bar{\te}f^{\ast}.
\end{eqnarray}
Noncommuting superfields $\widehat{V}, \widehat{\La}, \widehat{\bar{\La}}$
can be written in the same form in terms of their
components. At the first order in $\Te^{\mu \nu}$
let us denote the noncommutative fields as 
$\widehat{V}=V+V_{(1)},\ 
\widehat{\La}=\La+\La_{(1)},\ 
\widehat{\bar{\La}}=\bar{\La}+{\bar{\La}}_{(1)}$
and plug them into 
 the definition
(\ref{def1}). This will yield some  equations for 
component fields by matching the same $\te$ order terms.
In fact, the  equations including only  components of the superfields 
$\La$ and $\bar{\La}$ are
\beqa
\beta_{(1)} - \beta_{(1)}^{\ast}&=&0 ,\lb{s1s}\\
f_{(1)}=f_{(1)}^{\ast}=\kappa_{(1)}=\bar{\kappa}_{(1)}&=&0 . \lb{s2s}
\eeqa
Moreover, there are the equations
\beqa
A_{(1)\mu}(V_i + \kd V_i) -A_{(1)\mu}(V_i) -
\del_{\mu}\beta &=& -\Te^{\nu \rho}\del_{\nu}A_{\mu}\del_
{\rho}\beta , \\
\la_{(1)}(V_i + \kd V_i) - \la_{(1)}(V_i) &=& -
\Te^{\nu \rho}\del_{\nu}\la\del_{\rho}\beta , \\
\bar{\la}_{(1)}(V_i + \kd V_i) -
\bar{\la}_{(1)}(V_i) &=& - \Te^{\nu \rho}\del_{\nu}\bar{\la}\del_{\rho}
\beta ,\\
D_{(1)}(V_i + \kd V_i) - D_{(1)}(V_i) &=&
-\Te^{\nu \rho}\del_{\nu}D\del_{\rho}\beta , \lb{ed1}
\eeqa
where $V_i$ denotes the component fields.

Obviously, one can write  (\ref{def1}) in terms
of a general vector superfield instead of choosing the Wess--Zumino
gauge (\ref{Vx}), which would have 
drastically changed the equations for component fields.
However, we prefer to choose $V$ as (\ref{Vx}),
so that, we deal with the equations 
(\ref{s1s})--(\ref{ed1}) as defining 
supersymmetric Seiberg--Witten map.

 One can solve these  equations and get the noncommutative
fields in terms of the ordinary ones at the first order
in $\Te^{\mu\nu}$ as
\beqa
\widehat{A}_{\mu}&=& A_\mu -\fr{1}{2}\Te^{\nu \rho}(A_{\nu}\del_{\rho}
A_{\mu}+A_{\nu}F_{\rho\mu}) \lb{s1}, \\
\widehat{\la} & =& \la - \Te^{\nu \rho}\del_{\nu}\la A_{\rho} ,\lb{s2}\\
\widehat{\bar{\la}} &=& \bar{\la}- \Te^{\nu \rho}\del_{\nu}
\bar{\la} A_{\rho} , \\
\widehat{D}&=& D- \Te^{\nu \rho}\del_{\nu}D A_{\rho} \lb{s3s}.
\eeqa
(\ref{s1}) and (\ref{s2}) are also found in \cite{pss}
considering deformations of supersymmetric Yang--Mills theory
while preserving supersymmetry.

We also should define $\widehat{\psi}$ which is the noncommutative
component field resembling $\psi$
needed to define  a parent action 
to obtain duality transformation.
We define supersymmetric Seiberg--Witten map
of $\widehat{\psi}$ as
\beqa
\widehat{\psi}&=&\psi - \Te^{\nu \rho}\del_\nu \psi A_{\rho} \lb{pssw1}\\
\widehat{\bar{\psi}}&=& \bar{\psi}
- \Te^{\nu \rho}\del_\nu \bar{\psi} A_{\rho} \lb{pssw2},
\eeqa
which are consistent with (\ref{spsi}).

\section{Duals of noncommutative supersymmetric  $U(1)$ gauge theory}

Noncommutative generalization of supersymmetric $U(1)$ gauge 
theory\cite{ter} can be written in terms of the so called
 noncommuting
component fields, although they satisfy the usual
(anti)commutation relations, by the star product (\ref{star}) as  
\be
\lb{SNC}
S_{NC}=\fr{1}{2g^2}\int 
d^{4}x[-\fr{1}{2}\widehat{F}^{\mu\nu}\widehat{F}_{\mu\nu}
-i\widehat{\bar{\la}}\bar{\si}^{\mu}\widehat{D}_{\mu} \ast 
\widehat{\la}
-i\widehat{\la}\si^\mu \widehat{D}_{\mu} \ast\widehat{\bar{\la}}
+\widehat{D} \widehat{D}] ,
\ee
where $\widehat{D}_\mu \ast \widehat{\la}=\del_\mu\widehat{\la}
+i(\widehat{A}_\mu \ast \widehat{\la} -
\widehat{\la} \ast \widehat{A}_\mu ) $
and the noncommutative field strength is
$\widehat{F}_{\mu\nu}
=\del_\mu \widehat{A}_\nu -\del_\nu \widehat{A}_\mu 
+i(\widehat{A}_\mu \ast \widehat{A}_\nu -\widehat{A}_\nu \ast 
\widehat{A}_\mu ).$
It is invariant under the 
supersymmetry transformations given by the fermionic constant
spinor parameter $\xi$ as
\beqa
\kd_\xi\widehat{A}_\mu&=& i \xi\si^\mu\widehat{\bar{\la}}
+i \bar{\xi}\bar{\si}^\mu\widehat{\la}, \lb{SUP1}\\
\kd_\xi\widehat{\la}&=&\si^{\mu\nu}\xi\widehat{F}_{\mu\nu}
+i \xi\widehat{D}, \\
\kd_\xi\widehat{D}&=&\bar{\xi}\bar{\si}^\mu\widehat{D}_\mu 
\ast \widehat{\la}
- \xi\si^\mu\widehat{D}_\mu\ast \widehat{\bar{\la}}. \lb{SUP3}
\eeqa

Making use of the generalization of Seiberg--Witten map
to the supersymmetric case
(\ref{s1})--(\ref{s3s})
we write, up to the first order in $\Te ,$
the action of noncommutative supersymmetric U(1) gauge theory
(\ref{SNC})
 in terms of the ordinary component fields as
\beqa
S_{NC} [F,\la,D,\Te ] & = &\int 
d^{4}x\{-\fr{1}{4g^{2}}(F^{\mu\nu}F_{\mu\nu}+
2\Te^{\mu\nu}F_{\nu\ro}F^{\ro\si}F_{\si\mu}-\fr{1}{2}
\Te^{\mu\nu}F_{\nu\mu}F_{\ro\si}F^{\si\ro}) \nonumber \\
&&-\fr{i}{g^2}[\fr{1}{2}\bar{\la}\bar{\si}^{\mu}\del_{\mu}\la
+\Te^{\mu\nu}(\fr{1}{4}\bar{\la}\bar{\si}^{\ro}
\del_{\ro}\la F_{\mu\nu}+\fr{1}{2}\bar{\la}\bar{\si}^{\ro}
\del_\mu\la F_{\nu\ro}) \nonumber \\
&&
+ \fr{1}{2}\la \si^\mu \del_\mu \bar{\la}
+\Te^{\mu\nu}(\fr{1}{4}\la\si^{\ro}
\del_{\ro}\bar{\la}F_{\mu\nu}+\fr{1}{2}\la\si^{\ro}
\del_\mu\bar{\la} F_{\nu\ro})] \nonumber \\
&&
+\fr{1}{2g^2}(D^{2}+\fr{1}{2}\Te^{\mu\nu}D^{2}F_{\mu\nu})\}
\lb{ncaf}
\eeqa
Obviously,
when we write this  action we set the surface terms to zero
while 
performing  partial integrals.
The same action was also obtained in \cite{pw}
using a completely different approach.

Supersymmetry transformations 
which leave (\ref{ncaf}) invariant
can be read from (\ref{SUP1})--(\ref{SUP3}) as
\beqa
\kd_\xi A_\mu&=&i \xi\si_\mu \bar{\la} +
i\bar{\xi}\bar{\si}_\mu \la - i\Te^{\rho \ka}(\xi\si_\rho\bar{\la}+
\bar{\xi}\bar{\si}_\rho\la)(\fr{1}{2}F_{\ka\mu}
+\fr{1}{2}\del_\ka A_\mu) \nonumber \\
&& -i\Te^{\rho \ka} \fr{1}{2}(\xi\si_\rho
\del_\mu\bar{\la}+\bar{\xi}
\bar{\si}_\rho\del_\mu\la)A_\ka ,\\
\kd_\xi\la&=&\si^{\mu\nu}\xi F_{\mu\nu}+ i\xi D
+\Te^{\rho \ka}\del_\rho \la (i \xi\si_\ka \bar{\la}+i\bar{\xi}
\bar{\si}_\ka\la) \nonumber \\
&& +\Te^{\rho \ka}i \si^{\mu\nu}\xi F_{\mu \rho} F_{\nu \ka}, \\
\kd_\xi D&=&\bar{\xi}\bar{\si}^\mu\del_\mu\la - \xi\si^\mu
\del_\mu\bar{\la}-i\Te^{\rho \ka}(\xi\si_\rho\bar{\la}+
\bar{\xi}\bar{\si}_\rho\la)\del_\ka D \nonumber \\
&&+\Te^{\rho \ka}\xi\si^\mu F_{\rho\mu}\del_\ka\bar{\la}
-\Te^{\rho \ka}\bar{\xi}\bar{\si}^\mu F_{\rho\mu}\del_\ka\la  .
\eeqa

We would like to generalize the parent actions of the 
ordinary supersymmetric gauge theory 
(\ref{sfdc}) and (\ref{pr0}) to the noncommutative 
case. To this aim let us first take
$\widehat{F}_{\mu\nu}$  complex and deal with 
\beqa
I_{oNC} & = & -\fr{1}{2g^2}\int d^{4}x[
\fr{1}{4}
\widehat{F}^{\mu\nu}\widehat{F}_{\mu\nu}
+\fr{i}{8}\ep^{\mu\nu\rho \si}
\widehat{F}_{\mu\nu}\widehat{F}_{\rho\si}
+\fr{1}{4}
\widehat{F}^{\dagger \mu\nu}\widehat{F}^\dagger_{ \mu\nu}
+\fr{i}{8}\ep^{\mu\nu\rho \si}
\widehat{F}^\dagger_{ \mu\nu}\widehat{F}^\dagger_{\rho\si}\nonumber \\
& & +i\widehat{\la}\si^{\mu}\widehat{D}_{\mu} \ast
\widehat{\bar{\psi}}
+i\widehat{\bar{\la}}\bar{\si}^{\mu}\widehat{D}_{\mu} \ast
\widehat{\psi} 
-\fr{1}{2}\widehat{D}^2
-\fr{1}{2}\widehat{D}^{\dagger 2} \lb{i2d}
 ]  .
\eeqa
It is possible to discuss  supersymmetry and
gauge transformations of (\ref{i2d}), however, it
is not needed for the purposes of this work. 

Although the transformations (\ref{s1})--(\ref{pssw2}) 
are derived for a real vector superfield,
we suppose that they
are also valid for complex fields. We 
 perform the transformations 
(\ref{s1})--(\ref{pssw2}) and their complex conjugates
to write (\ref{i2d}) as
\beqa
I_{oNC}[F, \la , \psi , D] & = & 
I_o[F, \la , \psi , D]  -\fr{\Te^{\mu\nu}}{g^2}\int d^4x [
\fr{1}{4}F^{\rho\si}F_{\rho\mu}F_{\nu\si}
+\fr{1}{16}F_{\mu\nu}
F^{\rho\si}F_{\rho\si} \nonumber \\
& & 
+\fr{i}{8}\ep^{\la\ka\rho\si}
F_{\la\ka}F_{\rho\mu}F_{\nu\si}
+\fr{i}{32}\ep^{\la\ka\rho\si}
F_{\mu\nu}F_{\la\ka}F_{\rho\si}) \nonumber \\
&&\fr{i}{4}\la\si^{\rho}\del_{\rho}\bar{\psi}F_{\mu\nu}
-\fr{i}{2}\la\si^{\rho}\del_{\nu}\bar{\psi}F_{\mu\rho}
-\fr{1}{4}F_{\mu\nu}D^{2}+ {\rm c.c.} ], \lb{cncp}
\eeqa
where $I_o$ is defined in (\ref{iod}).
We define the
parent action 
\be
I_P=I_{oNC}[\ti{F}, \la , \psi , \ti{D}] +I_l,
\ee
where $I_l$ is given in (\ref{ild}).
We would like to emphasize that $\ti{F}_{\mu\nu}$
is not a field strength but a complex, antisymmetric
field.
When the  solutions of the
equations of motion with respect to dual fields 
(\ref{hdd1})--(\ref{spsi})
are
used in the parent action, it leads to
the noncommutative supersymmetric $U(1)$
gauge theory action (\ref{ncaf}).
However, when the equations of motion with respect to the
 fields 
$\ti{F}, \la , \psi , \ti{D}$ and their complex conjugates
are solved and used in the parent action
(\ref{cncp})
one finds
\be
\lb{doda}
I_{DNC}  = I_D+\fr{g^4}{4}\Te^{\mu\nu}\int d^4x
\ep^{\la\ka\rho\si}
[F_{D\la\ka}F_{D\rho\mu}F_{D\nu\si}
+\fr{1}{4}
F_{D\mu\nu}F_{D\la\ka}F_{D\rho\si}],
\ee
where $F_D$ is the field strength of $A_D.$
Obviously, we cannot define any duality symmetry
between (\ref{ncaf}) and  (\ref{doda}). The latter 
does not possess any contribution in terms
of the fields $\la, D$ at the first order in $\Te_{\mu\nu}.$

As the other possibility, let us
take 
$\widehat{F}_{\mu\nu}$ real
and deal with 
\be
S_{oNC}=\int d^{4}x[-\fr{1}{4g^{2}}\widehat{F}^{\mu\nu} 
\widehat{F}_{\mu\nu}-\fr{i}{2g^{2}}\widehat{\bar{\la}}
\bar{\si}^{\mu} \widehat{D}_\mu \ast \widehat{\psi} - \fr{i}{2g^{2}}
\widehat{\la} \si^{\mu} \widehat{D}_\mu \ast \widehat{\bar{\psi}}
+\fr{1}{2g^{2}}\widehat{D}\widehat{D}^\dag ]. \lb{S0D}
\ee
Through the  supersymmetric Seiberg--Witten map
(\ref{s1})--(\ref{pssw2}),
we  write the
 action (\ref{S0D}) as
\beqa
S_{oNC} [F, \la , \psi , D]&=& 
\int d^{4}x\{-\fr{1}{4g^{2}}(F^{\mu\nu}F_{\mu\nu}+
2\Te^{\mu\nu}F_{\nu\ro}F^{\ro\si}F_{\si\mu}-\fr{1}{2}
\Te^{\mu\nu}F_{\nu\mu}F_{\ro\si}F^{\si\ro}) \nonumber \\
&&-\fr{i}{2g^{2}}(\bar{\la}\bar{\si}^{\mu}\del_\mu \psi
+\Te^{\mu\nu}\bar{\la}\bar{\si}^{\ro} \del_\mu\psi F_{\nu\ro}
+\fr{1}{2}\Te^{\mu\nu}\bar{\la}\bar{\si}^{\ro}\del_\ro\psi F_{\mu\nu}) \nonumber \\
&&- \fr{i}{2g^{2}}(\la\si^{\mu}\del_\mu \bar{\psi}+\Te^{\mu\nu}\la\si^{\ro}
\del_\mu\bar{\psi} F_{\nu\ro}+\fr{1}{2}\Te^{\mu\nu}
\la\si^{\ro}\del_\ro\bar{\psi} F_{\mu\nu}) \nonumber \\
& & + \fr{1}{4g^{2}}[D^{2}+D^{\dag 2}+\fr{1}{2}\Te^{\mu\nu}(D^{2}
+D^{\dag 2})F_{\mu\nu})]\} .
\eeqa
Now, we define the parent action as
\be
\lb{slbo}
S_{P}= S_{oNC}[F_R, \la , \psi , \ti{D}] +  S_l  
\ee
where as before $F_{R\mu\nu}$ denotes an antisymmetric real field
and the Legendre transformation part $S_l$ is given in (\ref{sl}).

Equations of motion with respect to the dual fields
$A_D,\la_D, \bar{\la}, D_D$ are given as
before by (\ref{m1})--(\ref{m4}). 
Plugging their solutions 
into $S_{oNC}$ leads to the noncommutative supersymmetric
$U(1)$ gauge theory (\ref{ncaf}).

Equations of motion with respect to the other fields are
\beqa
 -\fr{1}{g^{2}}F_R^{\mu\nu}-\fr{1}{g^{2}}\Te^{\ro[\mu}F_R^{\nu]\si}
F_{R\si\ro}-\fr{1}{2g^{2}}\Te^{\ro\si}
F_{R\si[\mu}F_{R\nu]\ro}+\fr{1}{4g^{2}}
\Te^{\mu\nu}F_{R\ro\si}F_R^{\ro\si} \nonumber \\
+\fr{1}{2g^{2}}\Te^{\ro\si}F_{R\ro\si}F_R^{\mu\nu} 
- \fr{i}{2g^{2}}(\Te^{\ro\mu}\bar{\la}\bar{\si}^{\nu}-\Te^{\ro\nu}
\bar{\la}\bar{\si}^{\mu})\del_\ro\psi- \fr{i}{2g^{2}}\Te^{\mu\nu}
(\bar{\la}\bar{\si}^{\ro}\del_\ro\psi) \nonumber \\
-\fr{i}{2g^{2}}(\Te^{\ro\mu}\la\si^{\nu}-\Te^{\ro\nu}\la\si^{\mu})
\del_\ro\bar{\psi} - \fr{i}{2g^{2}}\Te^{\mu\nu}\la\si^{\ro}\del_\ro
\bar{\psi} \nonumber \\
+\fr{1}{4g^{2}}\Te^{\mu\nu}(\ti{D}^{2}+\ti{D}^{\dag 2})
-\vae^{\mu\nu\ro\si}\del_{\ro}A_{D\si} = 0 ,\\
 - \fr{i}{2g^{2}}\si^{\mu}\del_\mu \bar{\psi}-
\fr{i}{4g^{2}}\Te^{\mu\nu}\si^{\ro}\del_\ro\bar{\psi} F_{R\mu\nu}
- \fr{i}{2g^{2}}\Te^{\mu\nu}\si^{\ro}\del_\mu\bar{\psi} F_{R\nu\ro}
+\fr{1}{2}\si^{\mu}\del_{\mu}\bar{\la}_{D} =  0 ,\\
- \fr{i}{2g^{2}}\bar{\si}^{\mu}\del_\mu \psi
- \fr{i}{4g^{2}}\Te^{\mu\nu}\bar{\si}^{\ro}\del_\ro\psi F_{R\mu\nu}
- \fr{i}{2g^{2}}\Te^{\mu\nu}\bar{\si}^{\ro}\del_\mu\psi F_{R\nu\ro}
- \fr{1}{2}\bar{\si}^{\mu}\del_{\mu}\la_{D} =  0 ,\\
- \del_\mu\left[\fr{i}{2g^{2}}\bar{\la}\bar{\si}^{\mu}
- \fr{i}{4g^{2}}\Te^{\rho\nu}\bar{\la}\bar{\si}^{\mu}F_{R\rho\nu}
-\fr{i}{2g^{2}}\Te^{\mu\nu}\bar{\la}\bar{\si}^{\ro}F_{R\nu\ro}
- \fr{1}{2}\bar{\la}_{D}\bar{\si}_{\mu}\right] =  0, \\
\del_\mu\left[-\fr{i}{2g^{2}}\la\si^{\mu} - \fr{i}{4g^{2}}
\Te^{\ro\nu}\la\si^{\mu}F_{R\ro\nu} - \fr{i}{2g^{2}}\Te^{\mu\nu}
\la\si^{\ro}F_{R\nu\ro}+\fr{1}{2}\la_{D}\si^\mu\right]  =  0,\\
\fr{1}{2g^{2}}\ti{D}+\fr{1}{4g^{2}}\Te^{\mu\nu}\ti{D}F_{R\mu\nu}
+ \fr{i}{4}D_{D} =  0 ,\\
\fr{1}{2g^{2}}\ti{D}^\dag +\fr{1}{4g^{2}}\Te^{\mu\nu}
\ti{D}^\dag F_{R\mu\nu}-\fr{i}{4}D_{D} =  0 .
\eeqa
We solve these equations for $F_R, \psi, \la , \ti{D}$
and plug the solutions into 
(\ref{slbo}) to obtain the dual action 
\beqa
S_{NCD}&=&\int d^{4}x[- \fr{g^{2}}{4}(F_{D}^{\mu\nu}F_{D\mu\nu}
+2\ti{\Te}^{\mu\nu}F_{D\nu\ro}F_{D}^{\ro\si}F_{D\si\mu}
-\fr{1}{2}\ti{\Te}^{\mu\nu}F_{D\nu\mu}F_{D\ro\si}F^{D\si\ro}) \nonumber \\
&&-i g^{2}(\fr{1}{2}\la_{D}\si^{\mu}\del_{\mu}\bar{\la}_{D}
+\fr{1}{2}\bar{\la}_D \bar{\si}^\mu \del_\mu\la_D
+\fr{1}{4}\ti{\Te}^{\mu\nu}\la_{D}\si_{\mu}\del^\ro\bar{\la}_{D}
F_{D\ro\nu}) \lb{dns} \\
&&
+\fr{1}{4}\ti{\Te}^{\mu\nu}\bar{\la}_{D}\bar{\si}_{\mu}
\del^\ro \la_{D}
F_{D\ro\nu})  +\fr{1}{2}(D_{D}^{2}+
\fr{1}{2}\ti{\Te}^{\mu\nu}D_{D}^{2}F_{D\mu\nu}) \nonumber 
],
\eeqa
where 
\be
\lb{tt}
\ti{\Te}^{\mu \nu}\equiv g^2\ep^{\mu\nu\rho \si}\Te_{\rho\si},
\ee

When the fermionic and auxiliary fields 
$ \la_D,D_D$
set equal to zero one obtains the result of \cite{grs}:
There is a  duality
symmetry  under the replacement of $A^\mu$ with $A^\mu_D$
and $\Te_{\mu\nu}$ with $\ti{\Te}_{\mu\nu}.$ Unfortunately,
this symmetry accompanied by the replacement of
$\la, D$ with $ \la_D,D_D,$
cease to exist between the noncommutative
supersymmetric action
(\ref{ncaf}) and its dual (\ref{dns}).
Inspecting the terms which obstruct the duality
symmetry we can find actions in terms of
the component fields which possess this symmetry.
Let us  define the action
\be
\lb{sda}
{\Sigma}(\Te ,F,\la,\bar{\la},D)
=S_{NC} 
-\fr{i}{g^2}\int d^4x \Te^{\mu \nu} \left(
\la\si_\mu\del^\rho \bar{\la}+ \bar{\la} \bar{\si}_\mu
\del^\rho \la \right) F_{\rho \nu},
\ee
which can be obtained from the parent action
\be
\lb{sdpa}
{\Sigma}_P =S_P -\fr{i}{2g^2}\int d^4x \Te^{\mu \nu} \left(
\psi\si_\mu\del^\rho \bar{\la}+ \bar{\psi} \bar{\si}_\mu
\del^\rho \la 
+\la\si_\mu\del^\rho \bar{\psi}+ \bar{\la} \bar{\si}_\mu
\del^\rho \psi 
\right) F_{R\rho \nu},
\ee
when the solutions of 
equations of motion with respect to dual fields
$A_D, \la_D,D_D$ are plugged into it.
Now, the
dual theory which follows from (\ref{sdpa}) can be shown to be
\be
{\Sigma}_D
=g^4{\Sigma}(\ti{\Te} ,F_D,\la_D,\bar{\la}_D,D_D).
\ee
Therefore we conclude that the action (\ref{sda}) 
possesses the duality symmetry when the original fields
are substituted by the dual ones and the noncommutativity
parameter $\Te$ is replaced with $\ti{\Te}.$ However,
whether the action (\ref{sda}) is supersymmetric or not
is an open question. However, it is explicitly gauge invariant.

\vspace{1cm}

\noindent
{\bf Acknowledgment:} \"{O}.F. Dayi and B. Yapi\c{s}kan 
thank the Abdus  Salam International
Centre for Theoretical Physics High Energy Group where 
a part of this work is done.


\end{document}